\documentclass{article}
\usepackage{amsfonts}
\begin{document}

\title {Asymptotic  Liapunov Exponents Spectrum For An Extended
                        Chaotic Coupled Map Lattice}

\author{
D. Volchenkov
\footnote{Universit\"{a}t Bielefeld, Zentrum f\"{u}r
Interdisziplin\"{a}re Forschung, Wellenberg 1 D-33615 Bielefeld
(Germany); dima.volchenkov@uni-bielefeld.de}
and R. Lima
\footnote{Centre de Physique Theorique, CNRS, Luminy Case-907, 13288
Marseille (France); lima@univ-mrs.fr}}

\maketitle
\large

\begin{abstract}
The scaling hypothesis for the coupled chaotic map lattices (CML)
is formulated. Scaling properties of the CML in the regime of
{\it extensive} chaos observed numerically  before is justified
analytically. The asymptotic Liapunov exponents spectrum for
coupled  piece-wise linear chaotic maps is found.
\end{abstract}

\section{Introduction}

\subsection{The Liapunov exponents spectrum}

Partial differential equations describing continuous models and
real systems can be discretized into a system  of coupled maps
called as coupled map lattices (CML). In the last fifteen years, a
close attention has also been drawn to these models in virtue of
studies of the generic properties of spatiotemporal chaos,
\cite{1}-\cite{2}. Nevertheless, by this time, just a few rigorous
analytical results on the chaotic dynamics observed in CML
 have been achieved, \cite{BK}-\cite{BC}.

Chaotic behavior in extended systems is typically incoherent,
\cite{ChMann}. The problem of stability in lattice dynamical
systems has been extensively discussed in \cite{BuCa}. The
long-time limit along orbit of the rate of separation of points
in the tangent motion is given by the Liapunov exponents spectrum
\cite{CH},
\begin{equation}
\lambda_k=\lim_{n\to \infty}  \frac 1n
\sum_{t=0}^{n-1}\log \left|(DF^t)\right|_k,
\label{le}
\end{equation}
where $(DF^t)$ is the Jacobian matrix evaluated on the orbit
$F^t$. The notation $|\ldots|_k$ denotes the $k$-th eigenvalue of
the matrix $(DF),$ ordered in such a way that
$\left|(DF)\right|_{k-1}\geq\left|(DF)\right|_{k}\geq
\left|(DF)\right|_{k+1}.$ The spectrum (\ref{le}) characterizes the
stability of  motion in the phase space. If all exponents
$\lambda_k$ are negative the trajectory has an attractive  fixed
point. If the maximal value of the spectrum $\lambda_0=0$, the
attractor is a limit cycle, whereas positive exponents
correspond to diverging orbits.

A simple example of CML is given by the 1-dimensional chain
\begin{equation}
F(c_{x}^{t}) \equiv c_x^{t+1}=(1-2\varepsilon)f(c^t_x)+ {\varepsilon}
\left[f(c^t_{x-1})+f(c^t_{x+1})\right], \quad x\in {\mathbb Z},
\quad t\in {\mathbb Z_+},
\label{nnc}
\end{equation}
of loosely coupled ($\varepsilon < 1/4$) piecewise linear maps
\begin{equation}
f(c^{t+1}_x)= rc_x^t \quad \mathrm{mod}\quad  1,
\label{map}
\end{equation}
which is known to induce the chaotic behavior as the value of the
control parameter becomes  $r\geq 1$. Space-time chaos properties
in chains of weakly interacting hyperbolic maps has been
considered in \cite{PS}.

The Liapunov exponent of the solitary uncoupled map (\ref{map}) is
known exactly as a function of the control parameter $r,$
$\lambda_0=\log r.$ Furthermore, since the correspondent Jacobian
in (\ref{nnc}) is a $N\times N$ tridiagonal matrix, its
eigenvalues can be computed explicitly for any finite number of
sites $N$ provided the boundary conditions ($c_{-1}$ and
$c_{N+1}$) are known, \cite{Smith}. For example, chosen the  zero
valued  boundary conditions, $c_{-1}=c_{N+1}=0,$ one obtains the
following Liapunov exponents spectrum,
\begin{equation}
\lambda_k=\lambda_0+\log\left(1-2\varepsilon+2\varepsilon\cos
\frac{\pi k}{N+1}\right), \quad k=1\ldots N.
\label{lk}
\end{equation}
For any finite $N$,  the spectrum (\ref{lk}) is bounded from above
by the maximal value coinciding with the Liapunov exponent of
uncoupled maps $\lambda_{0}$. Either of the spectral values
$\lambda_k<\lambda_0$ depend both on the boundary conditions
chosen and the actual value of the coupling strength parameter
$\varepsilon.$ In the extended limit, $N\to \infty,$ the boundary
conditions can be assumed as an asymptote $\lim_{N\to \infty}
c_{\pm N}=0,$ and the spectrum (\ref{lk}) turns to a band. In
actual numerical simulations of the extended limit, the boundary
conditions are usually taken as periodic.

More realistic models for partial differential equations  refer
to more complicated coupled maps lattices for which the Liapunov
exponents spectra have not been found. Therefore, one is
interested in methods which can yield at least qualitative
information on them.

The Liapunov exponents spectra of CML in the extended limit,
$N\to\infty,$ is of special interest, \cite{ChMann}. Intuitively,
if the system is large enough, subsystems that are sufficiently
separated in space become decorrelated and so contribute to the
fractal dimension in proportion to their number.  This  phenomenon
usually called as an {\em extensive} chaos, \cite{HS}, is
numerically observed for lattice sizes larger than some threshold
value depending on the model, \cite{BC}.

For large lattices, the number of positive Liapunov exponents
$\lambda_k >0$ scales like the lattice size
\begin{equation}
N_{\lambda_k>0}\sim  N/{\cal U},
\label{scale}
\end{equation}
where $N$ is the volume of a large albeit finite piece $\mathbb{
L}$ of the lattice, and ${\cal U}$ is said to be the volume of a
typical statistically independent block. In accordance with
numerical simulations data, ${\cal U}$ depends substantially on
the particular model while the relation (\ref{scale}) is believed
to hold generally. The result (\ref{scale}) has  been observed
numerically for some simple coupled map lattices (see \cite{C}
and references therein for a review). Nevertheless, up to our
knowledge, the scaling (\ref{scale}) has not been justified
analytically, in particular, the  meaning of the statistically
independent block ${\cal U}$ is still vague.

Another fascinating phenomenon observed in the chaotic coupled
map lattices \cite{Hak}-\cite{ChMann} is a non-trivial collective
behavior ("hidden coherence", \cite{K}), where there is
apparently a unique nice invariant measure, \cite{MK}, but
nevertheless non-uniqueness of nice measures on the space-time
configurations, because  a two-cycle or more complicated dynamic
behavior occurs in time. The time periodic behavior of measures
has been discussed in \cite{LM} ("asymptotic periodicity").

The chaotic dynamics in CML lacks of mathematically rigorous
results. Nevertheless, some "physical" approaches borrowed from
statistical physics and quantum mechanics can be applied to the
theory  of chaotic spatially extended dynamical systems. An
essential problem of the "physical" point of view is that one
implies the nice invariant measure (or nice measures if the
system is subjected to a phase transition) is defined. In general,
coupled maps are not the case except the coupling parameter
$\varepsilon$ is not vanishingly small, \cite{BS}.

Our example of coupled maps chain (\ref{nnc}) has an invariant
measure defined as a Gibbs state even if $\varepsilon$ is not
small, \cite{Jprivate}. The non-trivial collective behavior
("hidden coherence") is not actually captured by the present
paper. At the same time, while computing the generalized entropy
$K_q$ and the partition function $\mathbf{Z}$, we shall perform
the summations $\sum_{\{\xi\}}$ over "all configurations"
${\{\xi\}}$. The summation technique that we use relies upon the
symmetry properties of the coupled chaotic maps chain and is
known as a {\it collective coordinates } method. Proceeding with
computations, we replace $\sum_{\{\xi\}}$ with an integration over
an infinite set of continuous real valued positive variables.
This stroke also implies that there is a number of nice measures
on the space-time configurations.

\subsection{The scaling hypothesis for coupled map lattices}

The {\it scaling} hypothesis in thermodynamics has been an
important milestone in formulation  of the  modern phase
transitions theory. It is based on two fundamental principles of
non-equilibrium thermodynamics:

i) while interesting in critical phenomena, one can  substitute a
true micro-model for a {\it fluctuation} one;

ii) in the  regime of strongly developed fluctuations, the
initial values are of no consequences for the system behavior.

For the Ising model, the scaling hypothesis put forward by
Kadanoff \cite{Kad} leans upon the "blocks" procedure, in which
the interactions of spins have been replaced for the interactions
of "spin blocks". It consists of two propositions:

j) the Hamiltonian of "spin blocks" interaction has just
 the same form as the Hamiltonian of  spin interaction with
 parameters changed;

jj) the dependence of parameters on the block size $L$ has a
scaling form as $L\to\infty.$

For the uncoupled maps having a "short range time memory",  the
scaling hypothesis has been propounded first in \cite{EP},
\cite{PP} and then used in \cite{GBP}.

We will follow a track by considering fluctuating effective
quantities analogous to the Liapunov exponents in finite time
segments, $n\gg 1.$ The "effective Liapunov exponent" $\lambda^n$
define the rate of separation of points in tangent motion in
finite time. They fluctuates around  the true Liapunov exponent
$\lambda $ having the fluctuation range $\Lambda$,  \cite{GBP}:
\begin{equation}\label{def1}
  \lambda=\lim_{n\to \infty}\int d{\!}\Lambda {\
  }\rho_n(\Lambda)\lambda^n,
\end{equation}
in which $\rho_n(\Lambda)$ is the probability distribution to
observe fluctuation $\Lambda$ in $n$ time steps. In the limit
$n\to\infty$, $\lambda^n$ form a micro-canonical ensemble
$\lim_{n\to\infty}\rho_n(\Lambda)=\delta(\Lambda).$

An example of  $\rho_n(\Lambda)$, one can consider the Fig.1
where  we have plotted the data of $5000$ numerical simulations
for the uncoupled solitary map (\ref{map}) at $r=1.6,$ for the
effective Liapunov exponent $\lambda^{n},$ $n=700.$ The graph
shows the probability distribution to observe deviations from the
true Liapunov exponent $\lambda_0\approx 0.47\ldots.$ When $n$
growths up, the bell-shaped profile becomes sharper.

 Being a product of $n$ Jacobi matrices, the effective rates
$\lambda^{n},$ for the maps with a short-range time memory, behave
like averages of $n$ random variables correlated only over short
time. Therefore, their values, for $n$ large,  do not depend on
the particular choice of initial conditions.

The scaling hypothesis formulated in \cite{EP} for such maps
claims on the {\it self-similarity } property for the
microscopical entropy, $S=\log \rho(\Lambda)$ in a long however
finite time string, $n>>1$, in the chaotic regime:
\begin{equation}
\label{sh}
\rho_n(\Lambda)\propto \exp[-S],\quad S=ng(\Lambda)
\end{equation}
where $ g(\Lambda) $ is a  "scaling function" (a "free energy")
pertinent to the map, \cite{EP}.

  In the Fig.2, we have pictured out the ratio of
scaling functions $ g_{n_1}/g_{n_2}$ via $\Lambda $ computed
numerically for the map (\ref{map}) in the sequences of $n_1=600$
and $n_2=1200$ time steps. The picture data show that there is an
interval of deviations $(\simeq[-0.1,0.1])$ within which the
scaling function $ g(\Lambda) $ is almost {\it universal} at
least in the first $1200$ time steps.

We note that in accordance with (\ref{sh}), given a smooth
function $F(\Lambda),$ the relevant observable is $\langle
F\rangle=\int d{\!}\Lambda {\ }
\rho(\Lambda)F(\Lambda).$ Since $n$ is a large number, one infers
 that  the main contribution to this integral comes from  the saddle-point
configuration $\Lambda_{*}$ such that $g(\Lambda_{*})$ is the
maximal $g$.

Indeed, this hypothesis breaks down for maps with strong long-time
correlations, when a minor change of initial conditions can stir
the behavior even in a very long time segments.

In the present paper, we  extend the scaling hypothesis  to the
case of coupled maps lattices. We  search for the statistics of
fluctuating quantities $\lambda^n_k$ analogous to the true
Liapunov exponents $\lambda_k$ in a long albeit finite time. We
consider the extended lattices ($N\to\infty$) of coupled maps with
a short range time memory. By the way, the boundary conditions
are of no consequences for the system behavior in "thermodynamic
limit".

Therefore, following \cite{Kad} and resting on the numerical
simulations data cited in \cite{C}, we propose that in the
chaotic regime the "free energy" function $g_\mathcal{U}(\Lambda)$
pertinent to the CML is universal up to the statistically
independent block $\mathcal{U},$ so that the probability to
observe a deviation of range $\Lambda=|\lambda^n_k-\lambda_k|$ in
the thermodynamic limit equals to
\begin{equation}
\label{sh1}
\rho(\Lambda)\propto \exp[-S],\quad
S=nN_\mathcal{U}g_\mathcal{U}(\Lambda).
\end{equation}
Furthermore, we propose that in the chaotic regime the scaling
function pertinent to the coupled maps lattice, $g_\mathcal{U},$
up to a change of parameters equals to that one for the solitary
uncoupled map, $g.$

In thermodynamic limit, according to (\ref{sh1}), the most
important contribution to observables comes from saddle-points
configurations $\Lambda_*$ (provided they exist) maximizing the
scaling function $g_\mathcal{U}.$

\subsection{Connections of a loosely coupled maps chain
 to a  quantum mechanical system with a
double-well potential}

Let us remark on a semblance between the statistics of effective
Liapunov exponents of the  simplest loosely coupled
($\varepsilon\ll 1$) maps chain (\ref{nnc}) in  thermodynamic
limit and of the quantum mechanical system with a double-well
potential in the infinite $\beta$ limit (i.e., in small
temperature limit, $T\to 0$) \cite{Zinn}.

 For $\beta\gg 1$, the main
contribution to the partition function of the quantum mechanical
problem comes from the {\it instanton} (or saddle-points)
configurations which renders the action stationary. In the limit
$\beta\to \infty,$ a linear combination of instanton solutions
plays a role. The relevant contributions do not correspond to
solutions of the classical equations of motion since these
equations are non-linear. However, in the limit when all
instantons are largely separated, such a combination also renders
the action almost stationary since the relevant corrections
coming from each instanton are exponentially small in the
separation.

To properly study the energy spectrum of the quantum mechanical
problem at $T\to 0$, first, one has to introduce a set of
collective variables (or coordinates) which describe fluctuations
in distances between the instantons, and, second, to fix out their
values in such a way that minimizes the variation of the action.

In the present paper, we are going to apply this method to
compute the asymptotic of a partition function $\mathbf{Z}$ over
the space of possible configurations $\{c^t_x\}$ generated by the
chain of coupled maps (\ref{nnc}).

In Fig. 3.1, we have plotted the results of $10000$ numerical
simulations for the chain of $N=8$ coupled maps (the value of
coupling strength parameter has been taken as $\varepsilon =0.02$)
supplied by the zero valued boundary conditions. This graph shows
the probability distribution profile to observe the effective
Liapunov exponents $\lambda^n$ in the interval $[0.44\ldots 0.48]$
in $n=1000$ iterations. One can see that almost all values
$\lambda^n$ drop around the true Liapunov exponents laying within
this interval, $\lambda_0\simeq 0.4700\ldots,$ $\lambda_1\simeq
0.4660\ldots,$ and $\lambda_2\simeq 0.4548\ldots$. The quota of
$\lambda^n$ arising  apart from the close vicinities of Liapunov
exponents $\lambda_i$ decreases very fast as $ n $ grows up.

In Fig. 3.2, we have drawn the distribution profile at $n=2500$
time steps. For $n\gg 1,$ the dispersion $\delta$ becomes much
smaller than the distances $\Delta\lambda_i$ between  peaks,
$\delta\ll \Delta\lambda_i.$ This means that fluctuations which
tend to change  distances between Liapunov exponents induce just a
vanishingly small variation of  $g$ at long time segments.

We note that  the thermodynamic  limit  when   fluctuations of
$\lambda^n_k$ are faded out, plays the role of the infinite
separation limit for instantons in the quantum well problem
mentioned above.

The plan of the paper follows. In the Sec. 2, leaning on the
thermodynamic formalism as applied to CML, we explain what kind
of information on the Liapunov exponents spectrum one could
obtain from the relevant partition function $\mathbf{Z}.$ Then,
in Sec. 3 and 4, we compute $\mathbf{Z}$  using  the scaling
hypothesis formulated in the previous subsection (1.2) and
following methods developed in quantum mechanics with respect to
the double-well potential problem \cite{Zinn}. Namely, in the Sec.
3, we compute the scaling function $g(\Lambda)$ (the free energy
of the microscopical interaction) for the map (\ref{map}) coupled
merely to the nearest neighbors. In the Sec. 4, we introduce the
block scaling function $g_\mathcal{U}(\Lambda)$ (the block
interaction) and then compute the asymptotic of $\mathbf{Z}$ in
thermodynamic limit utilizing the collective variables method. We
conclude in the Sec. 5.

\section{Thermodynamic formalism and the Liapunov exponents}

Thermodynamic formalism relies on the existence of a symbolic
representation for the dynamics. The idea  to study this
representation via Gibbs states for the $d+1$-dimensional system
goes back to \cite{S} and \cite{R}.

Let us suppose that one can split the phase space of CML $L\subset
{\mathbb Z^d}$ into  disjoint  boxes $b_i$ of size $a$. We shall
call as an {\it orbit} generated  by the coupled map in  site
$x_i$ the sequence of boxes, $\{b_1,\ldots b_j,\ldots b_n\},$ in
which the coupled map takes values at  consequent  time steps
$j=1\ldots n.$ We shall call as a {\it configuration} $\xi^n_L$
the set of orbits,
$$
\xi^n_L=\left\{ \{b_1,\ldots b_n\}_{x_i}, x_i\in L\right\},
$$
in the finite piece $L\subset {\mathbb Z}^d.$ The thermodynamic
formalism comes about by asking for a conditional probability
$P\left(\left.\xi^n_L\right| \xi^n_{{\mathbb Z^d}\setminus
L}\right) $ to observe the particular configuration $\xi^n_L$
given the configuration in the complement $\xi^n_{{\mathbb
Z^d}\setminus L}.$ This probability looks like an expectation in
the canonical ensemble,
\begin{equation}
P\left(\left.\xi^n_L\right| \xi^n_{{\mathbb Z^d}\setminus
L}\right) =\mathbf{Z}^{-1}\exp\left[ -\beta\Phi
\left(\left.\xi^n_L\right| \xi^n_{{\mathbb Z^d}\setminus L}\right)\right],
\label{P}
\end{equation}
in which the normalization factor (the partition function,
in terms borrowed from statistical physics) is
\begin{equation}
\mathbf{Z}(L;\xi^n_{{\mathbb Z^d}\setminus L}) =
\sum_{\{ \xi^n_L\}}\exp
\left[ -\beta\Phi
\left(\left.\xi^n_L\right| \xi^n_{{\mathbb Z^d}\setminus L}\right)\right]
\label{Pf}
\end{equation}
(the sum is taken over all configurations $\xi^n_L$ depending
upon possible configurations $\xi^n_{{\mathbb Z^d}\setminus L}$
in the complement). Here, $\beta>0$ is an inverse temperature
parameter representing a "heat bath" of coupling  maps.

The interaction potential $\Phi$ plays the role of a Hamiltonian
defined for the finite sublattice $L$ by
\begin{equation}
\Phi
\left(\left.\xi^n_L\right| \xi^n_{{\mathbb Z^d}\setminus L}\right)=
\sum_{n\in {\mathbb Z_+}}\log\left| \det (DF)
\left(\left.\xi^n_L\right| \xi^n_{{\mathbb Z^d}\setminus L}\right)
\right|.
\label{fi}
\end{equation}
For infinitely long time strings, it depends upon the
configuration $\xi_{\mathbb Z^d}$ in the whole lattice. The
Jacobi matrix  $(DF)$ can be found in the form $(DF)=B(I-W)$
where $B$ is the diagonal part of the matrix which provides a
point-wise contribution  from maps with no coupling, and $W$
comes from  coupling. The coupling  $W$ possesses the translation
invariance property. Therefore, the potential (\ref{fi}) can be
extended  naturally  to all subsets of the lattice $\mathbb{Z}^d$
by translation invariance.

For future convenience, we use the identity $\det \equiv \exp
\mathrm{Tr}\log $ to  rewrite the potential (\ref{fi}) in the following form
\begin{equation}
\Phi=
\sum_{n\in {\mathbb Z_+}} {\ }Tr {\ }\left[ \log (DF)
\left(\left.\xi^n_L\right| \xi^n_{{\mathbb Z^d}\setminus L}\right)
\right],
\label{fi1}
\end{equation}
which has been discussed extensively in \cite{BK}-\cite{JP}, and
\cite{MK}. Due to expansion $\log (DF) = \log B +\log (I-W)$, the
Hamiltonian (\ref{fi1}) can be divided in two folds:
$\Phi_L=\Phi_0+ U, $ where $\Phi_0$ is a "free" Hamiltonian
relevant to the uncoupled maps and the  coupling "interaction"
$U.$

Using the definition of effective Liapunov exponents,
$\lambda^n_k,$  one can express the partition function (\ref{Pf})
 as follows,
\begin{equation}
\mathbf{Z}=\lim_{n\to \infty} \sum_{\{ \lambda^n_k\}}\exp\left(n\beta
\mathrm{Tr}[\lambda^n_k]\right)= \exp\left(\beta\mathrm{Tr}[\lambda_k] \right)
\label{Pf1}
\end{equation}
where the sum $\sum_{\{ \lambda^n_k\}}$ is taken over the
ensemble of fluctuating quantities $\lambda^n_k.$ For the
uncoupled maps, the Eq. (\ref{Pf1}) yields $\lambda =
\lim_{n\to\infty}\log\mathbf{Z}/n\beta.$ However, for
coupled map lattices,  one can just obtain merely the sum of
positive Liapunov exponents  $\mathrm{Tr}[\lambda_k].$

It is essential to make use of the scaling hypothesis (Sec. 1.2)
while searching  for the Liapunov exponents. In particular,
 it implies (see Sec. 4) that
\begin{equation}\label{scale1}
\mathrm{Tr}[\lambda_k]\sim_{N\gg 1} N.
\end{equation}
Therefore, at least for large $N$, the values of the Liapunov
exponents $\lambda_N$ can be found by the simple subtraction,
$\lambda_N=\mathrm{Tr}[\lambda_{N-1}]-\mathrm{Tr}[\lambda_{N}].$
Furthermore, the maximal value in the spectrum, $\lambda_0$
equals to the Liapunov exponent of the uncoupled map and, at
least for the chain (\ref{nnc}), is also known.

However, since the relation (\ref{scale1}) is formally valid only
for $N\gg 1$, the spectral values $\lambda _{k},$ $k\sim O(1),$
are still uncertain. The numerical simulations of systems
demonstrating the extensive chaotic  behavior (i.e., scaling)
give an evidence (see for example, \cite{BC}) in favour of that
the threshold lattice size is typically $N \simeq 8.$

\section{Scaling function $g(\Lambda)$ for the map coupled merely to
the nearest neighbors}

The scaling function $g(\Lambda)$ plays a role of free energy in
the ensemble of fluctuating effective Liapunov exponents. In the
limit $n\to\infty,$ it can be related to the generalized entropies
\cite{EP}
 $K_q$  defined by
\begin{equation}
K_q(\Lambda)=\frac 1{q-1}\lim_{n\to \infty}\frac 1{n}
\log\sum_{\{\lambda^n\}}
\rho_n(\Lambda)^q,
\label{ge}
\end{equation}
via Legendre transforms, \cite{EP},
\begin{equation}\label{lt}
 K_q(\Lambda)=\frac 1{q-1}\left[\Lambda q-g(\Lambda)\right], \quad
g(\Lambda)=\chi(q)-q\frac{\partial}{\partial q}\chi(q), \quad
\chi(q)\equiv(q-1)K_q.
\end{equation}
It has been extensively  discussed in a literature
 that $\lim_{q\to 0}K_q$ and
$\lim_{q\to 1}K_q$ are the topological and metric entropies,
respectively. An example of computation of $g(\Lambda)$ has been
given in \cite{EP} for the simple uncoupled two-dimensional
"baker's" map.

Let us  choose the site ${\bf 0}$ as an origin and introduce a
point-wise Hamiltonian
\begin{equation}
\Phi^\varepsilon_n({\bf 0})=\lambda_\varepsilon {}_0+
 U^\varepsilon_n({\bf 0},{ V}_{\bf 0}),
\label{pwH}
\end{equation}
in which the interaction potential $U^\varepsilon_n({\bf 0},{
V}_{\bf 0}),$ and $V_{\mathbf{0}}$ is a nearest neighborhood of
${\bf 0}.$ In the present section, we suppose that there is no
interactions between $V_{\bf 0}$ and remainder of the lattice.

The free  point-wise Hamiltonian equals to the logarithm of
diagonal elements of the Jacobi matrix, $\lambda_\varepsilon
{}_0=\log\left(r(1-2\varepsilon)\right).$ The  potential
$U^\varepsilon_n({\bf 0},{ V}_{\bf 0})$ describes the
instantaneous fluctuations of the effective Liapunov exponent. It
comprises of a sum of contributions coming from the neighboring
sites ${\mathbf{y}}\in {V}_{\bf 0},$
$$
U^\varepsilon_n({\bf 0},{V}_{\bf 0})=\sum_{{\bf y}
\in{ V_{\bf 0}}}u^\varepsilon_n{}_{\bf 0}({\bf y}).
$$
The relevant point-wise generalized entropy takes the form:
\begin{equation}
K_q(\varpi_\varepsilon)=\frac 1{q-1}\lim_{n\to \infty}\frac 1{n}
\log\sum_{\{\lambda^n\}} \exp \left[\beta q \varpi_\varepsilon-\beta q\sum_{{\bf y}
\in{ V_{\bf 0}}}u^\varepsilon_n{}_{\bf 0}({\bf y})\right],
\label{gepw}
\end{equation}
in which $\varpi_\varepsilon\equiv\lambda_0-\lambda_\epsilon
{}_0,$ and $\sum_{\{\lambda^n\}}$ denotes the sum over all
possible orbits $\{\lambda^n\}.$ Note, that the first term in the
exponent, $\mu\equiv \exp\left[\beta q \varpi_\varepsilon\right]$
persists for any orbit. It plays a role of the "fugacity" of the
fluctuation "gas".

To perform the summation in (\ref{gepw}) properly, we note that
the site ${\bf 0}$ and its nearest neighbors $V_{\bf 0}$ form an
isolated system. Therefore, any fluctuation risen in ${\bf 0}$
can be represented as a vector in the $V$-dimensional  space of
particular contributions from the neighboring sites,
$\mathbb{U}\equiv\{u^\varepsilon_n{}_{\bf 0}({\bf y})\}$. Since
the entropy of closed system does not change in time, the sum
$\sum_{\{\lambda^n\}}$ can be interpreted as an integral over the
set of linear transformations $\mathbb{U}\to \mathbb{U}$ which
render the entropy (\ref{gepw}) unchanged.

The transformations we are interested in is apparently defined by
the set of positive parameters $\varphi^n_{\bf y}>0,$ such that
\begin{equation}
\beta=\sum_{{\bf y} \in {\cal V}_0}\varphi^n_y, {\ } \forall n.
\label{constrain}
\end{equation}
Choosing at each time step $V-1$ values $\{\varphi^n_y\}$  at
random, one generates a particular  orbit  $\{\lambda^n\}.$
Therefore, one can  replace the  sum over possible orbits in
(\ref{gepw}) with  integrations over the  real positive variables
$\varphi ^n_y,$
\begin{equation}
K_q(\varpi_\varepsilon)=\frac 1{q-1}\lim_{n\to \infty}\frac 1{n}
\log\sum_{n\in\mathbb{Z}_+} \left[
\frac 1n
\int_{\varphi_y^n>0}
\prod_{n\in {\mathbb Z_+}}
\prod_{{\bf y}\in {V}_{\bf 0}} d{\!}\varphi^n_y \times \right.
\label{Ephi}
\end{equation}
$$
\times \left. e^{nq\left[\beta  {\varpi_\varepsilon} -
1/n\sum_{{\bf y} \in {V}_{\bf 0}}\varphi^n_y
u^\varepsilon_n{}_{\bf 0} ({\bf y})\right]}
\cdot\delta\left(\beta-\sum_{{\bf y}\in {  V}_{\bf 0}}\varphi^n_y\right)
\right].
$$
The factor $1/n$ before the integral over $\varphi^n_y$ in
(\ref{Ephi}) arises because the result of integration has to be
invariant under a cyclic permutation of $\varphi^n_y-$variables in
time.

After the  integration over  $\varphi^n_y$ variables
\cite{delta}, in (\ref{Ephi}), one obtains
\begin{equation}
K_q (\varpi_\varepsilon) =
\frac 1{q-1} \lim_{n\to \infty}\frac 1{n}
\log\sum_{n\in\mathbb{Z}_+}
\frac 1{2  \pi n i}\int_{-\eta-i\infty}^{-\eta+i\infty}
\frac{ e^{\beta(s + n\varpi_\varepsilon)}}
{  \mathrm{P}_V^{(n)}(s,q)} d{\!}s,
\label{B}
\end{equation}
in which $\mathrm{P}_V^{(n)}(s,q)=\prod_{n\in {\bf Z_+}}
\prod_{{\bf y} \in {V}_{\bf 0}}\left(s-q
u^\varepsilon_n{}_{\bf 0}({\bf y})
\right)$ is a the polynomial of the order $V$. For piecewise linear maps,
one can proceed further since the slope of map is a piecewise
constant, and the values $u_n{}_{\bf 0}({\bf y})$ in the
polynomial $\mathrm{P}_V^{(n)}(s,q)$ does not depend on $n$.

The sum $\sum_{n\in \mathbb{Z_+}}$ therefore can be easily
performed,
\begin{equation}\label{nn}
K_q(\varpi_\varepsilon)=\frac 1{q-1}
\log\left[
\frac 1{2 \pi i \beta}\int_{-\eta-i\infty}^{-\eta+i\infty}
d{\!}s  \beta e^{-\beta s}
\log w(s,\varpi_\varepsilon)
\right]
\end{equation}
where we have defined
$$
w(s,\varpi_\varepsilon)=1-
\frac{e^{-\beta \varpi_\varepsilon}}
{
\prod_{{\bf y} \in {V}_{\bf 0}}\left(s-q
u^\varepsilon_n{}_{\bf 0}({\bf y})
\right)
}.
$$
The remaining integration in (\ref{nn}) is performed by parts,
$$
K_q(\varpi_\varepsilon) =\frac 1{q-1}
\log\left[
\frac 1{2 \pi i \beta}\int_{-\eta-i\infty}^{-\eta+i\infty}
d{\!}s {\ } \beta e^{-\beta
s}\frac{w'(s,\varpi_\varepsilon)}{w(s,\varpi_\varepsilon)}\right].
$$
The contour can be deformed  in the half-plane $\Re (s)\leq 0$
 to enclose the poles of the integrand
given by the solutions of the equation
\begin{equation}
{\prod_{{\bf y} \in { V}_{\bf 0}}
\left(s-q u^\varepsilon_n{}_{\bf 0}({\bf y})
\right)
}=e^{-\beta \varpi_\varepsilon}.
\label{eqq}
\end{equation}
The solutions of the equation (\ref{eqq}) can be easily found,
\begin{equation}
s=qu^\varepsilon_n{}_{\bf 0} +\exp[-\frac {\beta}{V}
\varpi_\varepsilon].
\label{s}
\end{equation}
 Then, the residue theorem yields:
\begin{equation}
K_q(\varpi_\varepsilon)=\frac {\beta}{q-1}\left[ q
\frac{\varpi_\varepsilon}{V} + qu^\varepsilon_n{}_{\bf 0}-
e^{-\frac {\beta}{V}
\varpi_\varepsilon}
\right].
\label{peX}
\end{equation}
Using the Legendre transforms (\ref{lt}), one obtains  the
relevant scaling function,
\begin{equation}
g(\varpi_\varepsilon)=\beta e^{-\frac {\beta}{V}
\varpi_\varepsilon}.
\label{gpw}
\end{equation}
The scaling function $g(\Lambda)$ (\ref{gpw}) has a typical form
pertinent to one-instanton contributions as they appear in
different problems in quantum mechanics. We also note the relation
between the fugacity of fluctuation gas $\mu$ and $g(\Lambda),$
$\mu(\Lambda)=g^V(\Lambda)$.

\section{The partition function in the thermodynamic limit}

In accordance to the scaling hypothesis, while proceeding  to the
interactions between blocks, the Hamiltonian of the interaction
has to be the same as the microscopic one, except the values of
parameters. The general scaling  in the system (i.e., the fact
that the dependence of parameters on the block size $L$ has a
scaling form as $L\to\infty$) then follows as a natural
consequence from the latter preposition \cite{Wil}.

Proceeding from (\ref{gpw}), for the statistically independent
block $\mathcal{U}$ (which constructive definition we give below),
one obtains the relevant interaction,
\begin{equation}\label{bif}
  g_\mathcal{U}(\Lambda)=\exp\left[-\frac{\beta}{\mathcal{U}}\Lambda\right].
\end{equation}
It differs from (\ref{gpw}) just on the volume $\mathcal{U}.$

Let us consider a finite piece $\mathbb{L}$ of the lattice
$\mathbb{Z}^d$. We assume that  $\mathbb{L}$ is large enough to
comprise of many statistically independent blocks $\mathcal{U}.$
For each site $x\in \mathbb{L},$ we assign an inverse temperature
parameter $\theta_x>0.$ These parameters is to be subjected to two
obvious constrains expressing the idea of statistical
independency of the block from the rest of lattice:
\begin{enumerate}
\item
The statistically independent block has to be in thermodynamical
equilibrium with the remainder of lattice. Therefore,
\begin{equation}
\beta=\sum_{s=1}^{N_\mathcal{U}}\theta_{x_s}
\label{ThE1}
\end{equation}
where the summation is going over the sites belonging to the same
block $\mathcal{U}.$
\item
All parts of the sublattice $\mathbb{L}$ should be in
thermodynamic equilibrium with respect to each other,
\begin{equation}
\frac {\beta}{\mathcal{U}}=\frac
1{N_{\mathcal{U}}}\sum_{s=1}^{N}\theta_{x_s},
\label{ThE2}
\end{equation}
in which $\mathcal{U}$ denotes the number of lattice sites
belonging  to one statistically independent block, $N$ is a total
number of sites in $\mathbb{L},$ and $N_{\mathcal{U}}$ is a total
number of statistically independent blocks in $\mathbb{L}.$
\end{enumerate}
Pursuant to (\ref{sh1}), the block-wise partition function has a
scaling form,
\begin{equation}\label{Bpf}
  \mathbf{Z}=\sum_{\{\ \mathcal{U}\}}
   \exp\left[-\beta N_\mathcal{U}g_\mathcal{U}(\Lambda)\right]
\end{equation}
where the summation is performed over all statistically
independent (in the sense of  constrains 1.) and 2.) )
collections of sites $\{\mathcal{U}\}$. To compute this sum in
the thermodynamic limit ($N_{\mathcal{U}}\to \infty$), we employ
the collective variables method again.

In terms of the $\theta_x$-variables, the block-wise scaling
function reads as
$$
g_\mathcal{U}(\Lambda)=\exp\left[-\frac{\Lambda}{N_{\mathcal{U}}}
\sum_{s=1}^{N}\theta_{x_s}
\right],
$$
and therefore,
\begin{equation}\label{Bpf1}
  \mathbf{Z}=\sum_{k=1}^{N_{\mathcal{U}}}\left[
  \frac 1k \int_{\theta_{x_s}>0}\prod_{s=1}^{k} d\theta_{x_{s}} {\ }
   \exp\left[-\beta N_\mathcal{U}e^{-\frac{\Lambda}{N_{\mathcal{U}}}
\sum_{s=1}^{N}\theta_{x_s}}\right]
\delta\left(\beta- \sum_{s=1}^{N_\mathcal{U}}\theta_{x_s} \right)
\right]
\end{equation}
where the factor $1/k$ in front of the integral is risen due to
cyclic permutation of $\theta$-variables. We use the integral
representation for the $\delta-$function (\cite{delta}) and
introduce the new auxiliary variable $\alpha$ such that
\begin{equation}
\mathbf{Z}=\sum_{k=1}^{N_{\mathcal{U}}}
\left[
\frac 1{2\pi i}\int^{-\eta+i\infty}_{-\eta-i\infty}
d\alpha {\ }  e^{\beta \alpha}\sum_{k=1}^{N_{\mathcal{U}}} \frac
1{k} \left[J(\alpha)\right]^k
\right],
\label{Bpf2}
\end{equation}
with the definition:
$$
J(\alpha)=\int_0^{\infty}\exp\left[ \alpha \theta+
\beta N_{\mathcal{U}} e^{\theta \Lambda/{N_{\cal U}}}
\right]
{\ } d\theta.
$$
By giving to $\alpha$ a small negative real part, we have insured
the convergence of the  integral $J(\alpha)$. By means of the
change of variables: $\tau \equiv \beta N_{ \mathcal{ U}}
e^{-\theta \varpi_{\varepsilon}/{N_{\mathcal{U}}}},$ one obtains,
$$
J(\alpha)=\int_0^{\beta N_{ \mathcal{U}}} \frac{d\tau}{\tau}
\left(
\frac{\beta N_{\mathcal{U}}}{\tau}\right)
^{\frac{-{N_{\mathcal{U}}}\alpha}{\Lambda}}e^{-\tau}.
$$
In thermodynamic limit $N_{\mathcal{U}}\to \infty,$ this integral
converges to the $\Gamma-$function,
$$
J(\alpha)\sim (\beta N_{\mathcal{U}})^ {\frac{{N_{\mathcal{
U}}}\alpha}{\Lambda}}
\Gamma\left(
\frac{{N_{\cal U}}\alpha}{\Lambda}
\right).
$$
The sum over $k$ in (\ref{Bpf2}) can be performed easily:
\begin{equation}
\mathbf{Z}
= \left[
\frac 1{2\pi i}\int^{-\eta+i\infty}_{-\eta-i\infty}
d\alpha {\ } e^{\beta \alpha}
\log
\left[1-(\beta N_{\cal U})^
{\frac{{N_{\cal U}}\alpha}{\Lambda}}\Gamma\left(
\frac{{N_{\cal U}}\alpha}{\Lambda}\right) \right]
\right].
\label{Bpf3}
\end{equation}
The  integral in (\ref{Bpf3}) can be computed by parts. Due to
asymptotic behavior of the $\Gamma-$function, the  integral
converges, and the contour of integration can be deformed to
enclose the poles of the integrand in the half-plane
$Re(\alpha)\leq 0.$ These poles correspond to the solutions of the
equation:
\begin{equation}
 (\beta N_{\cal U})^
{\frac{{N_{\cal U}}\alpha}{\Lambda}}=
\Gamma\left(
\frac{{N_{\cal U}}\alpha}{\Lambda}
\right).
\label{eq}
\end{equation}
In the limit $N_{\cal U}\to \infty$, solutions of the equation
(\ref{eq}) are close to the poles of $\Gamma-$function, i.e. to
negative integers.

Since the coupling strength parameter $\varepsilon$ is small, one
can taylor the $\Gamma-$function in powers of $\varepsilon$ to
find out the approximate solutions of (\ref{eq}). In the first
order in $\varepsilon$, this procedure gives:
$$
\frac{\alpha_\ell N_{\cal U}}{-2\varepsilon}= - \ell, \quad \ell=1\ldots N.
$$
Finally, one arrives at the spectrum of possible values
\begin{equation}
\alpha_\ell=\frac{2\varepsilon \ell}{N_\mathcal{U}}.
\label{final}
\end{equation}
The asymptotic result for the  spectrum of Liapunov exponents
follows from (\ref{final}),
\begin{equation}
\lambda_{\ell} \simeq_{
N\to \infty, {\ }
 \varepsilon \to 0} \lambda_0 -2\varepsilon
+  \frac{2\varepsilon}{N}\ell +  \mathrm{O}(\varepsilon^2),
\quad
\ell=1\ldots N.
\label{res}
\end{equation}
The spectrum (\ref{res}) is consistent with the relation
(\ref{scale}) observed numerically for wide variety of CML. One
can see that for $N\gg 1$ and $\varepsilon$ small it resembles
closely the spectrum (\ref{lk}).

\section{Discussion and Conclusions}

In the present paper, we have proposed and analyzed the
statistical theory of fluctuations of the effective Liapunov
exponents which correspond to the true Liapunov exponents in
finite time segments.

We have formulated the scaling hypothesis for the CML and
justified the general scaling behavior (\ref{scale}) of the CML
in the extensive chaotic regime.

Due to presence of large parameters ($n$ and $N$), fluctuations
of the effective Liapunov exponents can be well described by the
saddle-point approximation.

In the framework of thermodynamical formalism in CML, we have
computed the scaling function $g(\Lambda)$ for the $1-$dimensional
chain of coupled piece-wise linear maps and demonstrated that it
has a form typical for the $1-$instanton contribution in the
quantum mechanical problem with a double-well potential.

The basic technical point of scaling hypothesis is that the
block-wise Hamiltonian appears to have the same form as the
micro-interaction. As a direct consequence of this assumption,
one arrives at the following scaling properties of the Liapunov
exponents spectrum $\lambda_k$, for large lattices, $N\gg 1$:
$$
N_{\lambda_k>0}\sim N/\mathcal{U}, \quad \mathrm{and} \quad
\mathrm{Tr}(\lambda_k)\sim N.
$$
They allow to compute the asymptote of the Liapunov exponents
spectrum for $N\gg 1.$

We should note that the developed technique can be also applied to
the  multidimensional coupled piecewise linear maps. However, it
probably fails in case of non-linear map lattices. Nevertheless,
since most of  non-linear maps can be approached by  piecewise
linear ones, we believe that our technique would also  give good
approximations to more complicated loosely  coupled non-linear
map lattices.

\section{Acknowledgment}

The authors are grateful to B. Fernandez, M. Jiang, and V.
Afraimovich for useful and encouraging discussions.

This work has been supported by the project "The Sciences of
Complexity: From Mathematics to technology to a Sustainable World"
of the Universit\"{a}t Bielefeld, Zentrum f\"{u}r
Interdisziplin\"{a}re Forschung (ZiF).

One of the authors (D.V.) has been supported by the CIES (France)
and by the Alexander von Humboldt Foundation (Germany).

\newpage

\begin{center}

CAPTION FOR FIGURES

\end{center}

 \textbf{Figure 1.}

The probability to observe deviations of $\lambda^{n},$ $n=700,$
from the true Liapunov exponent $\lambda_0\approx 0.47\ldots$
($5000$ simulations).

\textbf{Figure 2.}

  The ratio of
scaling functions $ g_{n_1}/g_{n_2}$ via the deviation range
$\Lambda $ computed  in the sequences of $n_1=600$ and $n_2=1200$
time steps.

\textbf{Figure 3.1}

The probability distribution profile to observe the effective
Liapunov exponents $\lambda^n$ in the interval $[0.36\ldots
0.46]$ in $n=1000$ iterations ($10^4$ simulations, the chain of
$N=8$ coupled maps, the coupling strength parameter, $\varepsilon
=0.02$).

\textbf{Figure 3.2}

The probability distribution profile to observe the effective
Liapunov exponents $\lambda^n$ in the interval $[0.36\ldots
0.46]$ in $n=2500$ iterations ($10^4$ simulations, the chain of
$N=8$ coupled maps, the coupling strength parameter, $\varepsilon
=0.02$).


\begin{thebibliography}{99}
\bibitem{delta}

To factorize the integral over the $\varphi^n_y$  variables in
(\ref{Ephi}), we introduce an integral representation for the
$\delta-$function:
$$
\delta\left(\beta-\sum_{y\in {\cal V}}\varphi^n_y\right)=
\frac 1{2\pi i}\int_{-\eta-i\infty}^{-\eta+i\infty}
d{\!}s   \exp\left[-s\left(\beta-\sum_{y\in {
V}}\varphi^n_y\right)\right],
\quad \eta>0.
$$
\end{thebibliography}
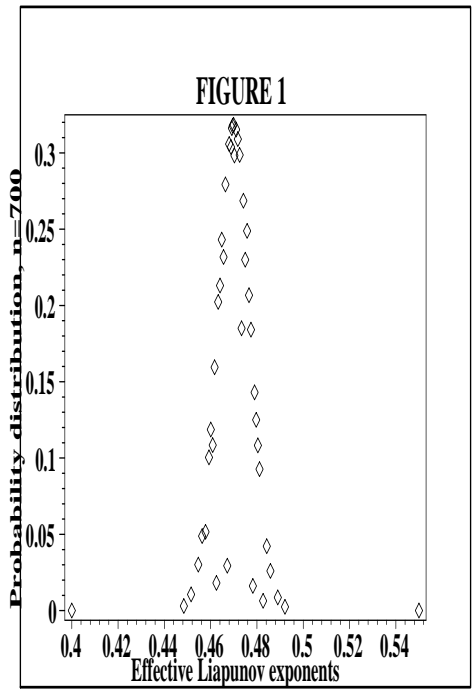
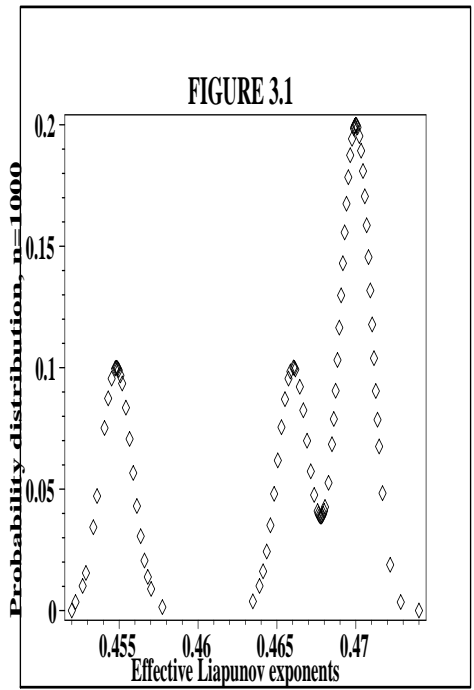
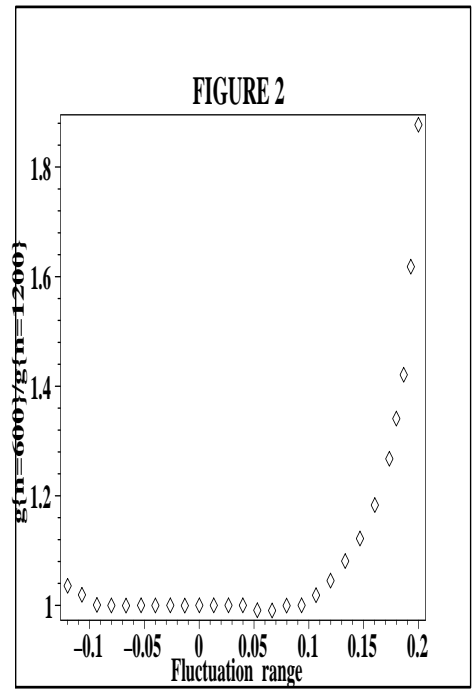
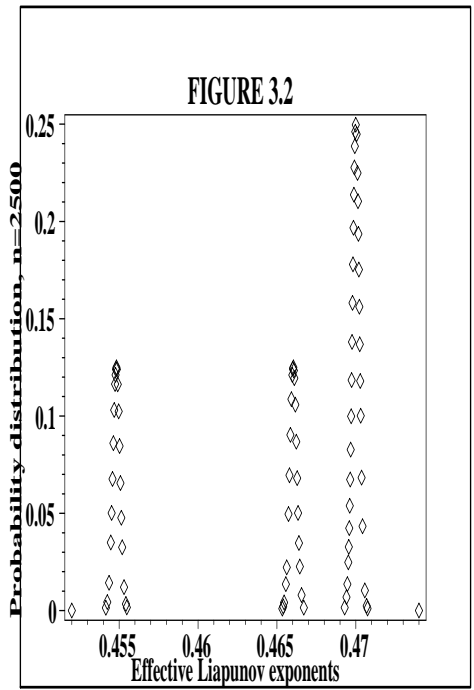
\end{document}